\documentclass[12pt,a4paper]{article}
\usepackage{times, natbib}
\usepackage{authblk}    
\usepackage{makeidx}    
\usepackage{graphicx}        
\usepackage{amsmath}
\usepackage{amsthm}
\usepackage[colorlinks=true, allcolors=blue]{hyperref}
\usepackage{booktabs}

\usepackage{amsmath,amsfonts,amssymb}
\usepackage{bbm}
\usepackage{bm}
\usepackage[ruled, norelsize, vlined]{algorithm2e}

\SetKwInput{kwInit}{Initialization}

\newcommand{\mailadd}{\href{mailto:me@somewhere.com}{\vspace*{0.0cm}}} 

\newcommand{\R}{ {\mathbb{R}} }
\newcommand{\E}{ {\mathbb{E}} }
\newcommand{\Var}{ {\text{var}} }
\newcommand{\new}{ {\textsc{new}} }

\newcommand{\diag}{ {\text{diag}} }

\newcommand{\br}{ {\bf r} }
\newcommand{\bt}{ {\bf t} }

\newcommand{\bx}{ {\bf x} }

\newcommand{\bS}{ {\bf S} }

\newcommand{\by}{ {\bf y} }
\newcommand{\bX}{ {\bf X} }

\newcommand{\bv}{ {\bf v} }

\newcommand{\bI}{ {\bf I} }
\newcommand{\bQ}{ {\bf Q} }
\newcommand{\bV}{ {\bf V} }

\newcommand{\bw}{ {\bf w} }

\newcommand{\bK}{ {\bf K} }

\newcommand{\bLambda}{ {\boldsymbol \Lambda} }

\newcommand{\bbeta}{ {\boldsymbol \beta} }
\newcommand{\bmu}{ {\boldsymbol \mu} }

\newcommand{\bSigma}{ {\boldsymbol \Sigma} }

\newcommand{\bOmega}{ {\boldsymbol \Omega} }
\newcommand{\bomega}{ {\boldsymbol \omega} }
\newcommand{\balpha}{ {\boldsymbol \alpha} }
\newcommand{\bxi}{ {\boldsymbol \xi} }

\setcitestyle{numbers} 
\setcitestyle{square}

\providecommand{\keywords}[1]
{
   \small	
  \hspace{1cm} \textbf{\textit{Keywords:}} #1
}

 \makeatletter
 \renewcommand{\section}{\@startsection
  {section}{1}{0pt}{24pt}{12pt}{\bf \large}}
 
 \makeatother

\title{\textbf{Efficient expectation propagation for posterior approximation in high-dimensional probit models}
}

\author[a]{Augusto Fasano}
\author[b]{Niccolò Anceschi}
\author[c]{Beatrice Franzolini}
\author[a,d]{Giovanni Rebaudo}
\affil[a]{\small Collegio Carlo Alberto, Turin, IT; \mailadd{\texttt{augusto.fasano@carloalberto.org}}}
\affil[b]{\small Duke University, Durham, USA;
\mailadd{\texttt{niccolo.anceschi@duke.edu}}}
\affil[c]{\small Agency for Science, Technology and Research (A*STAR), Singapore, SG; \mailadd{\texttt{beatricef@sics.a-star.edu.sg}}}
\affil[d]{\small University of Turin, Turin, IT;
\mailadd{\texttt{giovanni.rebaudo@unito.it}}}

\date{}
\begin{document}
\maketitle

\begin{abstract}
Bayesian binary regression is a prosperous area of research due to the computational challenges encountered by currently available methods either for high-dimensional settings or large datasets, or both.
In the present work, we focus on the expectation propagation (\textsc{ep}) approximation of the posterior distribution in Bayesian probit regression under a multivariate Gaussian prior distribution.
Adapting more general derivations in Anceschi et al.\ (2023), we show how to leverage results on the extended multivariate skew-normal distribution to derive an efficient implementation of the \textsc{ep} routine having a per-iteration cost that scales linearly in the number of covariates.
This makes \textsc{ep} computationally feasible also in challenging high-dimensional settings, as shown in a detailed simulation study.\\
\end{abstract}

\newpage
\section{Introduction and literature review}
\label{sec:1}
The past few years have seen florid research in Bayesian inference for the probit model \cite{durante2019,fasano2022scalable} as well as its extensions to dynamic \cite{fasano2021closed,fasano2021variational} and multinomial \cite{fasano2022class,fasano2022bayesian,loaiza2022fast} settings and beyond \cite{anceschi2023unified,loaiza2022fast2}.
This has been driven, among others, by computational challenges that may arise in high-dimensional settings.
See \cite{chopin2017leave} for an excellent review of Bayesian computations for binary regression.
Here, we focus on the expectation propagation (\textsc{ep}) approximation of the posterior of the Bayesian probit model
\begin{equation}
\label{eq:1}
\begin{split}
y_i\mid \bbeta &\overset{ind}{\sim} \textsc{Bern}\left(\Phi\left(\bx_i^\intercal \bbeta\right) \right),\quad i=1,\ldots,n,\\
\bbeta &\sim \textsc{N}_p(\boldsymbol{0},\nu^2 \bI_p),
\end {split}
\end{equation}
where $\bbeta\in \R^p$ is the unknown vector of parameters, $\bx_i\in \R^p$ is the covariate vector associated with observation $i$ and $\bI_p$ denotes the identity matrix of dimension~$p$.
$\Phi(t)$ denotes instead the cumulative distribution function of a standard Gaussian random variable evaluated at $t$.
Similarly, 
$\phi_p(\bt,\bS)$ will denote the density of a $p$-variate Gaussian random variable with mean $\boldsymbol{0}$ and covariance matrix $\bS$, evaluated at $\bt$.
\cite{durante2019} showed that the posterior distribution for model \eqref{eq:1} is a unified skew-normal (\textsc{sun}) and that, thanks to characterization properties of the \textsc{sun} family, one can obtain i.i.d.\ samples from it via a linear combination of $p$-variate Gaussian samples and $n$-variate truncated Gaussian samples.
As the computational bottleneck is represented by the truncated normal component, such i.i.d.\ sampler is well-suited for high-dimensional problems with small-to-moderate sample sizes but may become computationally hard for larger sample sizes.
To overcome such limitation, \cite{fasano2022scalable} developed a partially-factorized variational (\textsc{pfm-vb}) approximation of the posterior distribution which, for any fixed $n$, converges to the true posterior distribution as $p$ diverges.
Crucially, \textsc{pfm-vb} does not require dealing with any multivariate truncated Gaussian since the corresponding density component is replaced with a product of univariate truncated Gaussian densities, which do not represent a computational problem.
This approximation has a pre-processing cost of $\mathcal{O}(pn\cdot \min\{p,n\})$ and cost-per-iteration of $\mathcal{O}(n\cdot\min\{p,n\})$, making it computationally tractable also in large $p$ and large $n$ settings.
Empirically, the approximate posterior moments closely match the ones obtained via i.i.d.\ sampling for $p\ge 2n$.
The possible over-shrinkage of the posterior moments towards zero for smaller $p$ motivates the investigation of efficient implementations of other approximation techniques that may be more accurate in those settings, like \textsc{ep}, at the price of a higher computational cost.
Adapting more general results obtained for a broad class of models in \cite{anceschi2023unified}, we show how the \textsc{ep} routine for posterior inference under the multivariate Gaussian prior in \eqref{eq:1} can be implemented at per-iteration-cost of $\mathcal{O}(pn\cdot\min\{p,n\})$, which, although higher than the one of \textsc{pfm-vb}, improves over the cost $\mathcal{O}(p^2n)$ reported in \cite{chopin2017leave}, leading to sensible computational advantages and making \textsc{ep} computationally feasible also in settings with $p$ of the order of tens of thousands.
Considering the goodness of the \textsc{ep} approximation \cite{anceschi2023unified,chopin2017leave}, the possibility to extend the number of scenarios where it can be effectively implemented represents a major contribution to Bayesian binary regression computations.

\section{Expectation propagation for the probit model}
\label{sec:2}
In this section, we present an implementation of \textsc{ep} for the probit model \eqref{eq:1} which leverages results on multivariate extended skew-normal (\textsc{sn}) random variables (see \cite{azzalini2013skew}).
Calling $\by=(y_1,\ldots,y_n)$, in \textsc{ep} we approximate $p(\bbeta\mid\by)$ with $q(\bbeta) \propto \prod_{i=0}^{n}q_i(\bbeta)$, where $q_0(\bbeta),\ldots,q_n(\bbeta)$ are probability density functions and, in particular, $q_0(\bbeta) = p(\bbeta)$ and $q_i(\bbeta)\propto\exp\{-\frac{1}{2}\bbeta^\intercal\bQ_i\bbeta +\bbeta^\intercal\br_i\}$ for $i=1,\ldots,n$.
Hence, writing $q_0(\bbeta)\propto\exp\{-\frac{1}{2}\bbeta^\intercal\bQ_0\bbeta +\bbeta^\intercal\br_0\}$, with $\br_0 = \boldsymbol{0}$ and $\bQ_0=\nu^{-2}\bI_p$, we immediately note that $q(\bbeta) = \phi_p(\bbeta-\bQ^{-1}\br,\bQ^{-1})$, where $\br=\sum_{i=0}^n \br_i$, $\bQ = \sum_{i=0}^n\bQ_i$.

\textsc{ep} proceeds by updating each site $i=1,\ldots,n$ (we do not update the site of the prior), by iteratively matching the first two moments of the global approximation $q(\bbeta)$ and the hybrid distribution
\begin{equation}
\label{eq:2}
	h_i(\bbeta) \propto p(y_i\mid \bbeta) \prod_{j\ne i} q_j(\bbeta) =
	\Phi((2y_i-1)\bx_i^\intercal\bbeta)\prod_{j\ne i} q_j(\bbeta).
\end{equation}
To compute the moments of \eqref{eq:2}, instead of proceeding as \cite{chopin2017leave}, we can exploit the fact that some easy algebraic manipulations show that \eqref{eq:2} is the kernel of a multivariate extended skew-normal distribution $\textsc{sn}_p(\bxi_i,\bOmega_i,\balpha_i,\tau_i)$ (see \cite{azzalini2013skew}), with
\begin{equation*}
\begin{split}
    \bxi_i &= \bQ_{-i}^{-1}\br_{-i},\quad 
	\bOmega_i = \bQ_{-i}^{-1},\\
    \balpha_i &= (2y_i-1)\bomega_i\bx_i,\quad
	\tau_i = (2y_i-1)(1+\bx_i^\intercal\bOmega_i\bx_i)^{-1/2}\bx_i^\intercal\bxi_i,
\end{split}
\end{equation*}
where $\bQ_{-i}=\sum_{j\ne i} \bQ_j$, $\br_{-i}=\sum_{j\ne i}\br_j$ and $\bomega_i=\left[\text{diag}\left(\bOmega_i\right)\right]^{1/2}$.

\begin{algorithm}[b!]
 \caption{Probit \textsc{ep} - $\mathcal{O}(p^2 n)$ cost per iteration}
 \label{algo1}
\kwInit{\mbox{$\bQ^{-1} = \nu^2 \bI_p;\,\ \br=\boldsymbol{0}$; $\,\ k_i = 0$ and $m_i = 0$ for $i=1,\ldots,n$.}}
 \For{$\, t \,$ from $\, 1 \,$ until convergence $ $}{
 \For{$\, i \,$ from $\, 1 \,$ to $\, n \,$}{
 $\br_{-i} = \br - m_i \bx_i $\\ 
 $\bOmega_i = \bQ^{-1} + k_i/\left(1- k_i \bx_i^\intercal \bQ^{-1} \bx_i\right) \left(\bQ^{-1} \bx_i\right)\left(\bQ^{-1} \bx_i\right)^\intercal $ \\
 $s_i = (2y_i-1)(1+\bx_i^\intercal \bOmega_i \bx_i)^{-1/2}$ \\
 $\tau_i = s_i \bx_i^\intercal \bOmega_i \; \br_{-i} $\\[2pt]
 $k_i = -\zeta_2(\tau_i)/\left(1 + \bx_i^\intercal\bOmega_i\bx_i + \zeta_2(\tau_i)\bx_i^\intercal\bOmega_i\bx_i\right)$ \\[2pt]
 $m_i = \zeta_1(\tau_i) s_i + k_i (\bOmega_i\bx_i)^\intercal \br_{-i} + k_i \zeta_1(\tau_i) s_i \bx_i^\intercal \bOmega_i\bx_i $\\[2pt]
 $\br = \br_{-i} + m_i \bx_i $\\
 $\bQ^{-1}=\bOmega_i + \zeta_2(\tau_i) s_i^2(\bOmega_i\bx_i)(\bOmega_i\bx_i)^\intercal$
 }
 }
\KwOut{$ q(\bbeta)=\phi_p(\bbeta - \bQ^{-1}\br; \bQ^{-1}) $}
\end{algorithm}

After noticing this, exploiting formulae (5.71) and (5.72) in \cite{azzalini2013skew}, we can immediately obtain the first two moments of $h_i(\bbeta)$:
\begin{equation*}
\begin{split}
	\bmu_{h_i} &= \E_{h_i(\bbeta)} [\bbeta] = \bxi_i+\zeta_1(\tau_i) s_i\bOmega_i\bx_i\\
	\bSigma_{h_i} &= \Var_{h_i(\bbeta)} [\bbeta] = \bOmega_i + \zeta_2(\tau_i) s_i^2(\bOmega_i\bx_i)(\bOmega_i\bx_i)^\intercal,
\end{split}
\end{equation*}
where $s_i=(2y_i-1)(1+\bx_i^\intercal\bOmega_i\bx_i)^{-1/2}$, $\zeta_1(x) = \phi(x)/\Phi(x)$ and $\zeta_2(x)=-\zeta_1(x)^2-x\zeta_1(x)$.
Hence, when updating site $i$, the \textsc{ep} moment-matching condition implies that the updated quantities $\br_i^\new$ and $\bQ_i^\new$ must be such that
\begin{equation*}
	\begin{cases}
	\left(\bQ_{-i}+\bQ_i^\new\right)^{-1}(\br_{-i} +\br_i^\new) = \bmu_{h_i}\\
	\left(\bQ_{-i}+\bQ_i^\new\right)^{-1} = \bSigma_{h_i},
	\end{cases}
\end{equation*}
from which it immediately follows
\begin{equation*}
	\begin{cases}
	\br_i^\new = \left(\bQ_{-i}+\bQ_i^\new\right) \bmu_{h_i} - \br_{-i}\\
	\bQ_i^\new = \bSigma_{h_i}^{-1} - \bQ_{-i}.
	\end{cases}
\end{equation*}
The direct computation of $\bSigma_{h_i}^{-1}$ can be avoided since, by Woodbury's identity
\begin{equation*}
	\begin{split}
	\bQ_i^\new &
	=\bOmega_i^{-1} - \zeta_2(\tau_i) s_i^2 (1+\zeta_2(\tau_i) s_i^2\bx_i^\intercal\bOmega_i\bOmega_i^{-1}\bOmega_i\bx_i)^{-1}\bOmega_i^{-1}\bOmega_i \bx_i \bx_i^\intercal \bOmega_i\bOmega_i^{-1} -\bQ_{-i}\\
	& = -\zeta_2(\tau_i) s_i^2 (1+\zeta_2(\tau_i) s_i^2\bx_i^\intercal\bOmega_i\bx_i)^{-1}\bx_i \bx_i^\intercal
	= -(\zeta_2(\tau_i)^{-1} s_i^{-2} +\bx_i^\intercal\bOmega_i\bx_i)^{-1} \bx_i \bx_i^\intercal\\
	& = -\dfrac{\zeta_2(\tau_i)}{1 + \bx_i^\intercal\bOmega_i\bx_i + \zeta_2(\tau_i)\bx_i^\intercal\bOmega_i\bx_i} \bx_i \bx_i^\intercal
	= k_i^\new \bx_i \bx_i^\intercal,
	\end{split}
\end{equation*}
with $k_i^\new = -\zeta_2(\tau_i)/\left(1 + \bx_i^\intercal\bOmega_i\bx_i + \zeta_2(\tau_i)\bx_i^\intercal\bOmega_i\bx_i\right)$.
Moreover, 
\begin{equation*}
	\begin{split}
	\br_i^\new &= \bQ_{-i}\bmu_{h_i} + \bQ_i^\new\bmu_{h_i} - \br_{-i}= \bQ_{-i} \bQ_{-i}^{-1} \br_{-i} + \zeta_1(\tau_i) s_i \bQ_{-i}\bOmega_i\bx_i + \bQ_i^\new\bmu_{h_i} - \br_{-i}\\
	&=  \zeta_1(\tau_i) s_i \bx_i + \bQ_i^\new\bmu_{h_i}
	= \zeta_1(\tau_i) s_i \bx_i + k_i^\new \bx_i \bx_i^\intercal \bOmega_i \br_{-i} + k_i^\new \zeta_1(\tau_i) s_i \bx_i\bx_i^\intercal \bOmega_i\bx_i\\
	&= [\zeta_1(\tau_i) s_i + k_i^\new   (\bOmega_i\bx_i)^\intercal \br_{-i} + k_i^\new \zeta_1(\tau_i) s_i \bx_i^\intercal \bOmega_i\bx_i] \bx_i= m_i^\new \bx_i,
	\end{split}
\end{equation*}
where $m_i^\new = \zeta_1(\tau_i) s_i + k_i^\new   (\bOmega_i\bx_i)^\intercal \br_{-i} + k_i^\new \zeta_1(\tau_i) s_i \bx_i^\intercal \bOmega_i\bx_i$.
Hence, we can implement \textsc{ep} by storing only the scalar quantities $k_i$ and $m_i$, $i=1,\ldots,n$. 
In practice, they are initialized to zero, so that the initial global approximation is the prior distribution.
Combining the above results with Woodbury's identity, we obtain
\begin{equation*}
 \bOmega_i  = \bQ_{-i}^{-1} = \left(\bQ - k_i \bx_i \bx_i^\intercal \right)^{-1}
    =  \bQ^{-1} + \dfrac{k_i}{1- k_i \bx_i^\intercal \bQ^{-1} \bx_i} \left(\bQ^{-1} \bx_i\right)\left(\bQ^{-1} \bx_i\right)^\intercal,
\end{equation*}
which can be computed avoiding explicit matrix inversions, since $\bQ^{-1}$ is known from the beginning.
Finally, the update of the inverse of the \textsc{ep} precision matrix is immediate as $(\bQ^\new)^{-1}=(\bQ_{-i}+\bQ_i^{\new})^{-1}=\bSigma_{h_i}$.

Putting it all together, we obtain the \textsc{ep} implementation in Algorithm~\ref{algo1}.
Its core part coincides with the \textsc{ep} derivations presented in \cite{chopin2017leave}, and implemented in the \texttt{EPprobit} function in the \texttt{R} package \texttt{EPGLM}. However, we arrived at it by exploiting results on \textsc{sn}s that leverage more general derivations for a broader class of models presented in \cite{anceschi2023unified}.
We also avoided the computation of the normalizing constants for the unnormalized densities, $Z_i$, $i=1,\ldots,n$, which can be used for the computation of the approximate marginal likelihood, since the approximated posterior moments can be computed also without them.
Algorithm~\ref{algo1} has per-iteration cost $\mathcal{O}(p^2n)$, which, although avoiding explicit $p\times p$ matrix inversions, might be impractical in high-dimensional settings.
Adapting more general results presented in \cite{anceschi2023unified}, we thus derive in Section 
in full detail an implementation of \textsc{ep} for the Bayesian probit model having per-iteration cost~$\mathcal{O}(pn^2)$.

\section[Efficient expectation propagation for large p settings]{Efficient expectation propagation for large $p$ settings}
\label{sec:3}
The crucial part to obtain a per-iteration-cost that is linear in $p$ is to note that we can avoid handling $p\times p$ matrices, as, by close inspection of Algorithm \ref{algo1}, the whole \textsc{ep} routine can be written by working out directly the updates of the $p$-dimensional vectors $\bw_i = \bOmega_i \bx_i = \bQ_{-i}^{-1}\bx_i$ and $\bv_i = \bQ^{-1}\bx_i$, $i=1,\ldots,n$.
As for the former, we have
\begin{equation*}
	\begin{split}
	\bw_i &= \bQ_{-i}^{-1}\bx_i = (\bQ-\bQ_i)^{-1} \bx_i= \bQ^{-1}\bx_i + (1-k_i\bx_i^\intercal\bQ^{-1}\bx_i)^{-1} k_i (\bQ^{-1}\bx_i)(\bQ^{-1}\bx_i)^\intercal\bx_i\\
	&=\bv_i+ k_i (1-k_i\bx_i^\intercal\bv_i)^{-1} \bv_i\bv_i^\intercal \bx_i
	= \big[1+ (1-k_i\bx_i^\intercal\bv_i)^{-1} (k_i\bx_i^\intercal \bv_i)\big] \bv_i
	= d_i \bv_i,
	\end{split}
\end{equation*}
where $d_i=(1-k_i\bx_i^\intercal\bv_i)^{-1}$.
As for the $\bv_i$'s, each time a site $i$ is updated $\bQ^{-1}$ changes and thus all the $\bv_j$'s, $j=1,\ldots,n$, should be modified accordingly as
\begin{equation*}
\begin{split}
\bv_j^\new &= (\bQ^{\new})^{-1} \bx_j = (\bQ - \bQ_i+\bQ_i^\new)^{-1}\bx_j
= [\bQ + (k_i^\new - k_i) \bx_i \bx_i^\intercal]^{-1}\bx_j\\
&= [\bQ^{-1} - (k_i^\new - k_i) [1+(k_i^\new - k_i)\bx_i^\intercal\bQ^{-1}\bx_i ]^{-1} \bQ^{-1}\bx_i\bx_i^{\intercal}\bQ^{-1} ]\bx_j\\
&= \bQ^{-1}\bx_j-[(k_i^\new - k_i)^{-1} +\bx_i^\intercal\bv_i ]^{-1} \bv_i\bx_i^\intercal\bv_j
= \bv_j - c_i (\bx_i^\intercal\bv_j) \bv_i,
\end{split}
\end{equation*}
where $c_i=(k_i^\new - k_i)/ (1 + (k_i^\new - k_i)\bx_i^\intercal\bv_i)$.
Instead of cycling over $j$, these updates can be performed in block by defining a $p \times n$ matrix $\bV=[\bv_1,\bv_2,\dots,\bv_n]$.
Accordingly, $\bV^\new = \bV - c_i \bv_i \bx_i^\intercal \bV$. 
This operation is the most expensive per site update, being of order $\mathcal{O}(p n)$.
Accordingly, each \textsc{ep} iteration has cost $\mathcal{O}(p n^2)$.
Contrarily to Algorithm~\ref{algo1}, once the procedure has reached convergence we still need to calculate the inverse of the global precision matrix $\bQ^{-1}$.
The explicit calculation can be avoided as follows, obtaining a post-processing cost of $\mathcal{O}(p^2 n)$.
First, $\bQ = \bQ_0 + \sum_{i=1}^n k_i\bx_i \bx_i^\intercal = \nu^{-2} \bI_p + \bX^\intercal \bK \bX$ with $\bX=(\bx_1,\ldots,\bx_n)^\intercal$ and $\bK = \diag(k_1,\ldots,k_n)$.
Calling $\bLambda = (\bI_n+ \nu^2\bK\bX \bX^\intercal)^{-1}$, so that, by Woodbury's identity, $\bQ^{-1}=\nu^2 \bI_p - \nu^4 \bX^\intercal \bLambda \bK \bX$, one obtains that
$\bV = \bQ^{-1} \bX^\intercal
 = \nu^2 \bX^\intercal\big[\bI_n - \nu^2 \bLambda\bK \bX \bX^\intercal\big]=\nu^2 \bX^\intercal\bLambda\big[\bLambda^{-1} - \nu^2 \bK \bX \bX^\intercal\big]=\nu^2 \bX^\intercal\bLambda$ and thus $\bQ^{-1}=\nu^2\bI_p-\nu^2\bV\bK\bX$.
Notice that, if the interest is only in approximate posterior means and variances, this expression for $\bQ^{-1}$ allows doing it at reduced post-processing cost of $\mathcal{O}(p n)$.
The whole routine is summarized in Algorithm \ref{algo2}.
\begin{algorithm}[t]
    \label{algo2}
    \caption{Efficient probit \textsc{ep} for large $p$ - $\mathcal{O}(pn^2)$ cost per iteration}
 \kwInit{\mbox{$\br=\boldsymbol{0}$; $\,\ k_i = 0$ and $m_i = 0$ for $i=1,\ldots,n$; $\,\ \bV=\left[\bv_1,\ldots,\bv_n \right]=\nu^2\bX^\intercal$.}} 
 \For{$\, t \,$ from $\, 1 \,$ until convergence $ $}{
 \For{$\, i \,$ from $\, 1 \,$ to $\, n\,$}{
 $\bw_i = (1-k_i \bx_i^\intercal \bv_i)^{-1} \bv_i $ \\
 $\br_{-i} = \br - m_i \bx_i $\\[2pt]
 $s_i = (2y_i-1)(1+\bx_i^\intercal \bw_i)^{-1/2}$ \\
 $\tau_i = s_i \bw_i^\intercal \; \br_{-i} $\\[2pt]
 $k_i^\new = -\zeta_2(\tau_i)/\left(1 + \bx_i^\intercal\bw_i + \zeta_2(\tau_i)\bx_i^\intercal\bw_i\right)$\\[2pt]
 $m_i = \zeta_1(\tau_i) s_i + k_i^\new \bw_i^\intercal \br_{-i} + k_i^\new \zeta_1(\tau_i) s_i \bx_i^\intercal \bw_i $\\[2pt]
 $k_i = k_i^{\text{new}}$\\
 $\br = \br_{-i} + m_i \bx_i $ \\
 $\bV = \bV - \bv_i \left[ (k_i^\new - k_i)/ \left(1 + (k_i^\new - k_i) \bx_i^\intercal \bv_i\right) \right] \bx_i^\intercal \bV$
 }
 }
$\bQ^{-1}=\nu^2\bI_p-\nu^2\bV\bK\bX $\\[2pt]
\KwOut{$ q(\bbeta)=\phi_p(\bbeta - \bQ^{-1}\br; \bQ^{-1}) $}
\end{algorithm}

\section{Simulation study}
\label{sec:4}
\begin{figure}
    \centering
    \includegraphics[width=\linewidth,height=0.33\textheight]{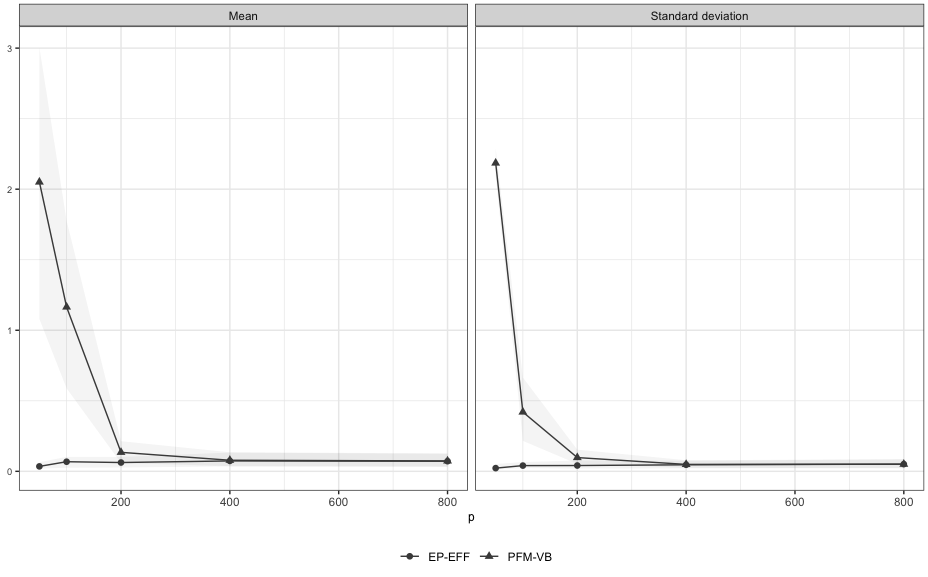}
    \caption{For varying $p$, median absolute difference between the $p$ posterior means and standard deviations resulting from $2000$ i.i.d.\ samples and the ones arising from \textsc{ep-eff} and \textsc{pfm-vb} for probit regression with $n=100$ and $\nu^2=25$.
    Grey areas denote the first and third quartiles.
    }
    \label{fig:1}
\end{figure}

We conclude with a simulation study where probit regression is applied to multiple simulated datasets, with $n = 100$ and $p = 50, 100, 200, 400$ and $800$.
We investigate the performances of \textsc{ep} when the efficient implementations presented in Algorithm~\ref{algo1} and Algorithm~\ref{algo2} are used when $p<n$ and $p\ge n$, respectively. 
Such implementation, denoted \textsc{ep-eff} in the following, is compared with \textsc{pfm-vb} in terms of running time and quality of the approximation.
The latter is measured by the median absolute difference between the approximate posterior means and standard deviations and the ones computed via $2000$ i.i.d.\ samples, for $\nu^2=25$.
The moderate sample size is taken so that the i.i.d.\ sampler is computationally efficient, but the approximate methods could be used in more challenging settings, as in all scenarios they both give almost immediate outputs.
To show the computational gains with respect to standard \textsc{ep} implementations, we also compare the running time needed to obtain the \textsc{ep} approximation with the \texttt{R} function \texttt{EPprobit} from the package \texttt{EPGLM}, which implements the \textsc{ep} derivations reported in \cite{chopin2017leave}.
As it emerges from Table \ref{table1}, \textsc{ep-eff} leads to a dramatic reduction of the computational effort with respect to the standard \texttt{EPprobit} in high dimensions. 
This results in a drop of the running time by more than three orders of magnitude in the setting $p=800$, with a computational gain increasing with $p$, as expected.
The \textsc{ep-eff} running times, although generally much lower than the ones of \texttt{EPprobit},
are still higher than the ones of \textsc{pfm-vb} in most cases.
Nevertheless, if one looks at the quality of the approximation of the two posterior moments in Figure \ref{fig:1}, \textsc{ep-eff} gives consistently accurate approximations across different dimensions of $p$, while \textsc{pfm-vb} gets similar accuracy for $p \gtrsim 2n$.
This shows the importance of developing efficient implementations for \textsc{ep} like the ones in this paper, so make it computationally feasible in challenging high-dimensional settings where routine implementations are impractical.
Code can be found at \href{https://github.com/augustofasano/EPprobit-SN}{https://github.com/augustofasano/EPprobit-SN}.
\begin{table*}[h]
\centering
\caption{\footnotesize{Running time, in seconds, to compute posterior means and standard deviations with the \textsc{ep} approximation as in Algorithms \ref{algo1} and \ref{algo2} (\textsc{ep-eff}), with the \textsc{ep} approximation computed via the \texttt{R} function \texttt{EPprobit} (\texttt{EPprobit}) and with the \textsc{pfm-vb} approximation (\textsc{pfm-vb}) for probit regression with $n=100$ and $\nu^2=25$.
}}
\label{table1}
\begin{tabular}[c]{ll|ccccc}
 \multicolumn{2}{l|}{}  &  \multicolumn{5}{c}{\textit{p}}   \\
\hline
 & \textit{Method} & 50 & 100 & 200 & 400 & 800 \\ 
   \hline
Running time (seconds) &\ \textsc{ep-eff} &\ 0.11 &\ 0.02 &\ 0.03 &\ 0.05 &\ 0.09 \\
&\ \texttt{EPprobit}\ &\ 0.07 &\ 0.42 &\ 3.18 &\ 24.36 &\ 140.24 \\
&\ \textsc{pfm-vb}\ &\ 0.11 &\ 0.06 &\ 0.01 &\ 0.01 &\ 0.01 \\

\hline
\end{tabular}
\end{table*}

\newpage
{\small
\paragraph{Acknowledgments}
The authors wish to thank D.\ Durante for carefully reading a preliminary version of this manuscript and providing insightful comments.}

\end{document}